\documentclass[5p,a4paper,sort&compress]{elsarticle}

\usepackage{amsmath}
\usepackage{pifont}
\usepackage{geometry}
\usepackage{fleqn}
\usepackage{multirow}
\usepackage{graphicx}
\usepackage{txfonts}
\usepackage{subfigure}
\usepackage{textcomp}
\usepackage{epsfig}
\usepackage{booktabs}
\usepackage{footnote}
\usepackage{ctable}
\usepackage{mathtools}
\usepackage{dcolumn}% Align table columns on decimal point
\usepackage{bm}% bold math

\newcommand\T{\rule{0pt}{2.6ex}}       % Top strut
\newcommand\B{\rule[-1.2ex]{0pt}{0pt}} % Bottom strut
\begin{document}

\title{Cross-shell states in $^{15}$C: a test for p-sd interactions }
%\title{Low-lying 2p-3h states in $^{15}$C: a test for p-sd effective interactions. }

% AUTHORS
\author[USC]{J. Lois-Fuentes}
\author[USC]{B. Fern\'{a}ndez-Dom\'{i}nguez\corref{mycorrespondingauthor}}
\cortext[mycorrespondingauthor]{Corresponding author}
\ead{beatriz.fernandez.dominguez@usc.es}
%\author[USC]{B. Fern\'{a}ndez-Dom\'{i}nguez}
\author[USC,LPC,KAON]{X. Pereira-L\'{o}pez}
\author[LPC]{F. Delaunay}
\author[Surrey]{W.N. Catford}
\author[Surrey,LPC]{A. Matta}
\author[LPC]{N.A. Orr}
\author[CEA,KU]{T. Duguet}
\author[CNS]{T. Otsuka}
\author[CEA]{V. Som\`a}
\author[GANIL]{O. Sorlin}
\author[NIH]{T. Suzuki}
%\author[CEA]{V. Som\`a}
%\author[CEA,KU]{T. Duguet}
%\author[GANIL]{O. Sorlin}
\author[LPC]{N.L. Achouri}
\author[IPNO]{M. Assi\'{e}}
\author[Bham]{S. Bailey}
\author[GANIL]{B. Bastin}
\author[IPNO]{Y. Blumenfeld}
\author[IFINHH]{R. Borcea}
\author[USC]{M. Caama\~{n}o}
\author[GANIL]{L. Caceres}
\author[GANIL]{E. Cl\'{e}ment}
\author[CEA]{A. Corsi}
\author[Bham]{N. Curtis}
\author[LPC]{Q. Deshayes}
\author[GANIL]{F. Farget}
\author[LNS]{M. Fisichella}
%\author[USC]{D. Fern\'{a}ndez-Fern\'{a}ndez}
\author[GANIL]{G. de France}
\author[IPNO]{S. Franchoo}
\author[Bham]{M. Freer}
\author[LPC]{J. Gibelin}
\author[CEA]{A. Gillibert}
\author[Regina]{G.F. Grinyer}
\author[IPNO]{F. Hammache}
\author[GANIL]{O. Kamalou}
\author[Surrey]{A. Knapton}
\author[Bham]{Tz. Kokalova}
\author[CEA]{V. Lapoux}
\author[IPNO]{B. Le Crom}
\author[LPC]{S. Leblond}
\author[LPC]{F.M. Marqu\'{e}s}
\author[IPNO]{P. Morfouace}
\author[GANIL]{J. Pancin}
\author[IPNO]{L. Perrot}
\author[GANIL]{J. Piot}
\author[CEA]{E. Pollacco}
\author[USC]{D. Ramos}
\author[USC]{D. Regueira-Castro}
\author[USC,GANIL]{C. Rodr\'{i}guez-Tajes}
\author[GANIL]{T. Roger}
\author[IFINHH]{F. Rotaru}
\author[LPC]{M. S\'{e}noville}
\author[IPNO]{N. de S\'{e}r\'{e}ville}
\author[Bham]{R. Smith\fnref{myfootnote1}}
\fntext[myfootnote1]{Present address:   Department of Engineering and Mathematics, Sheffield Hallam University, Howard Street, Sheffield, S1 1WB, United Kingdom.}
%\author[CEA]{V. Som\'a}
%\author[GANIL]{O. Sorlin}
\author[IFINHH]{M. Stanoiu}
\author[IPNO]{I. Stefan}
\author[GANIL]{C. Stodel}
\author[IPNO]{D. Suzuki}
%\author[CNS]{T. Suzuki}
\author[GANIL]{J.C. Thomas}
\author[Surrey]{N. Timofeyuk}
\author[GANIL,CEA]{M. Vandebrouck}
\author[Bham]{J. Walshe}
\author[Bham]{C. Wheldon}

% AFFILIATIONS

\address[USC]{IGFAE and Dpt. de F\'{i}sica de Part\'{i}culas, Univ. of Santiago de Compostela, E-15758, Santiago de Compostela, Spain}
\address[LPC]{LPC Caen, Normandie Universit\'e, ENSICAEN, UNICAEN, CNRS/IN2P3, Caen, France}
%\address[UTK]{Department of Physics and Astronomy, University of Tennessee, Knoxville, Tennessee 37996, USA}
%\address[York]{Department of Physics, University of York, Heslington, York YO10 5DD, United Kingdom}
\address[KAON]{Center for Exotic Nuclear Studies, Institute for Basic Science (IBS), Daejeon 34126, Republic of Korea}
\address[Surrey]{Department of Physics, University of Surrey, Guildford GU2 5XH, UK}
\address[CEA]{IRFU, CEA, Universit\'{e} Paris-Saclay, F-91191 Gif-sur-Yvette, France}
\address[KU]{KU Leuven, Instituut voor Kern- en Stralingsfysica, 3001 Leuven, Belgium}
\address[CNS]{CNS, University of Tokyo, 7-3-1 Hongo, Bunkyo-ku, Tokyo, Japan}
\address[GANIL]{GANIL, CEA/DRF-CNRS/IN2P3, Bd. Henri Becquerel, BP 55027, F-14076 Caen, France}
\address[NIH]{Department of Physics, College of Humanities and Sciences, Nihon University, Sakurajosui 3-25-40, Setagaya-ku, Tokyo, Japan}
%\address[CEA]{IRFU, CEA, Universit\'{e} Paris-Saclay, F-91191 Gif-sur-Yvette, France}
%\address[KU]{KU Leuven, Instituut voor Kern- en Stralingsfysica, 3001 Leuven, Belgium}
%\address[GANIL]{GANIL, CEA/DRF-CNRS/IN2P3, Bd. Henri Becquerel, BP 55027, F-14076 Caen, France}
%\address[IPNO]{IJCLab, Universit\'{e} Paris-Saclay, CNRS/IN2P3,91405 Orsay, France}
\address[IPNO]{Universit\'{e} Paris-Saclay, CNRS/IN2P3, IJCLab, 91405 Orsay, France}
\address[Bham]{School of Physics and Astronomy, University of Birmingham, Birmingham B15 2TT, UK}
%\address[GANIL]{GANIL, CEA/DRF-CNRS/IN2P3, Bd. Henri Becquerel, BP 55027, F-14076 Caen, France}
%\address[CEA]{IRFU, CEA, Universit\'{e} Paris-Saclay, F-91191 Gif-sur-Yvette, France}
\address[LNS]{INFN, Laboratori Nazionali del Sud, Via S. Sofia 44, Catania, Italy}
\address[Regina]{Department of Physics, University of Regina, Regina, SK S4S 0A2, Canada}
%\address[Sevilla]{Departamento de FAMN, Facultad de F\'{i}sica, Universidad de Sevilla, Apdo. 1065, E-41080 Sevilla, Spain}
%\address[Sevilla2]{Instituto Interuniversitario Carlos I de F\'{i}sica Te\'{o}rica y Computacional \(iC1\), Apdo. 1065, E-41080 Sevilla, Spain}
%\address[CNS]{CNS, University of Tokyo, 7-3-1 Hongo, Bunkyo-ku, Tokyo, Japan}
%\address[NIH]{Department of Physics, College of Humanities and Sciences, Nihon University, Sakurajosui 3-25-40, Setagaya-ku, Tokyo, Japan}
\address[IFINHH]{IFIN-HH, P. O. Box MG-6, 76900 Bucharest-Magurele, Romania}
%\address[KU]{KU Leuven, Instituut voor Kern- en Stralingsfysica, 3001 Leuven, Belgium}
%\curradr{Department of Engineering and Mathematics, Sheffield Hallam University, Howard Street, Sheffield, S1 1WB, United Kingdom}

\date{\today}

\begin{abstract}

  The low-lying structure of $^{15}$C has been investigated via the neutron-removal $^{16}$C$(d,t)$ reaction. Along with bound neutron sd-shell hole states, unbound p-shell hole states have been firmly confirmed. The excitation energies and the deduced spectroscopic factors of the cross-shell states are an important measure of the $[(p)^{-1}(sd)^{2}]$ neutron configurations in $^{15}$C. Our results show a very good agreement with shell-model calculations using the SFO-tls interaction for $^{15}$C. However, a modification of the $p$-$sd$ and $sd$-$sd$ monopole terms was applied in order to reproduce the $N=9$ isotone $^{17}$O. In addition, the excitation energies and spectroscopic factors have been compared to the first calculations of $^{15}$C with the \textit{ab initio} self-consistent Green's function method employing the NNLO$_{sat}$ interaction. The results show the sensitivity to the size of the $N=8$ shell gap and highlight the need of going beyond the current truncation scheme in the theory.
  %Reveals a signature of an increasing shell gap at N=8 as we decrase the isospin. The SFO-tls interaction agrees well.
%Negative parity states involving excitations across the N = 8 gap, are overestimated in the NNLO$_{sat}$ interaction. 

\end{abstract}


\maketitle

\section{Introduction}

The shell structure of nuclei is one of the fundamental cornerstones in nuclear physics. The evolution of the shell structure observed in exotic nuclei has led to a change of paradigm in our understanding of the nuclear force \cite{OtsukaRMP}. Nuclei far from the stability with large values of isospin are very sensitive to new aspects of nuclear forces. 

In particular, neutron-rich carbon isotopes have drawn a lot of attention due to their exotic properties: neutron halos \cite{Bazin,Fang}, clustering \cite{Freer}, and deformation \cite{Hamamoto}. Such light neutron-rich systems offer an ideal testing ground for theoretical models. On the one hand, phenomenological shell-model calculations in the $p$-$sd$ space with improved interactions, such as SFO-tls \cite{TSuzukiI,TSuzukiII} or YSOX \cite{YSOX} have been very successful in describing the structure of neutron-rich carbon isotopes \cite{Kim}. However, they fail at reproducing the energy levels of unnatural parity states across a given isotonic chain partly because of the lack of experimental data of cross-shell states in exotic nuclei. On the other hand, new interactions derived from chiral effective field theory \cite{chEFT_Epelbaum,chEFT_Machleidt} and rooted in quantum chromodynamics have made a significant progress in the description of light and medium-mass nuclei \cite{Ekstrom, VSoma3}. The low-lying spectroscopy of $^{15}$C obtained from the $(d,t)$ reaction is an ideal observable to constraint not only phenomenological interactions but also $NN$ and $3NF$ derived from chiral effective-field-theory. Excitation energies and spectroscopic factors determined from single-particle transfer reactions provide one of the most stringent tests to \textit{ab initio} wavefunctions derived from realistic nuclear forces.

We report here on the first investigation of the cross-shell states in $^{15}$C using the $^{16}$C$(d,t)$ single-neutron pick-up reaction. Assuming a normal ordering in $^{16}$C, protons occupy the $p$-shell while neutrons are placed in the $p$-$sd$-shell. Using $^{14}$C as a core, removing a neutron from the occupied orbitals (from the $^{16}$C ground state) will probe states with 1p configurations in $^{15}$C.  Additionally, extracting a neutron from the 0$p_{1/2}$ or 0$p_{3/2}$ orbitals gives rise to states with 2p-1h configurations across the 0$p_{1/2}$ and 1$s_{1/2}$-0$d_{5/2}$ orbits. Therefore, states with total spin and parity values of 1/2$^{-}$ and 3/2$^{-}$ and large single-particle component are signatures of particle-hole excitations and provide valuable information on the amplitude of the N=8 shell gap.  

Previous knowledge on the structure of $^{15}$C has been gathered over the past decades using a variety of experimental probes: single-neutron adding reactions $^{14}$C$(d,p)$ \cite{Cecil,Goss1,Goss2, Murillo,Kay}, two-neutron transfer $^{13}$C$(t,p)$ \cite{Truong} and $^{13}$C$(^{18}$O$,^{16}$O$)^{15}$C \cite{Cappuzzello}, Coulomb breakup \cite{UDatta}, $\beta$-delayed neutron decay of spin-polarized $^{15}$B \cite{Miyatake} and single-neutron knockout reactions from $^{16}$C \cite{SauvanPLB,Maddalena,Yamaguchi}. All these probes have provided extensive information, mainly on positive-parity states. However little is known about the negative-parity states. In particular, no p-hole states from single-neutron removal reactions are clearly established. Such states would be key for testing the shell-model interactions for the $p-sd$ model space, and potentially improving the description of light nuclei from one drip-line to the other.

\section{Experimental details}

\begin{figure}
\includegraphics[scale=0.28]{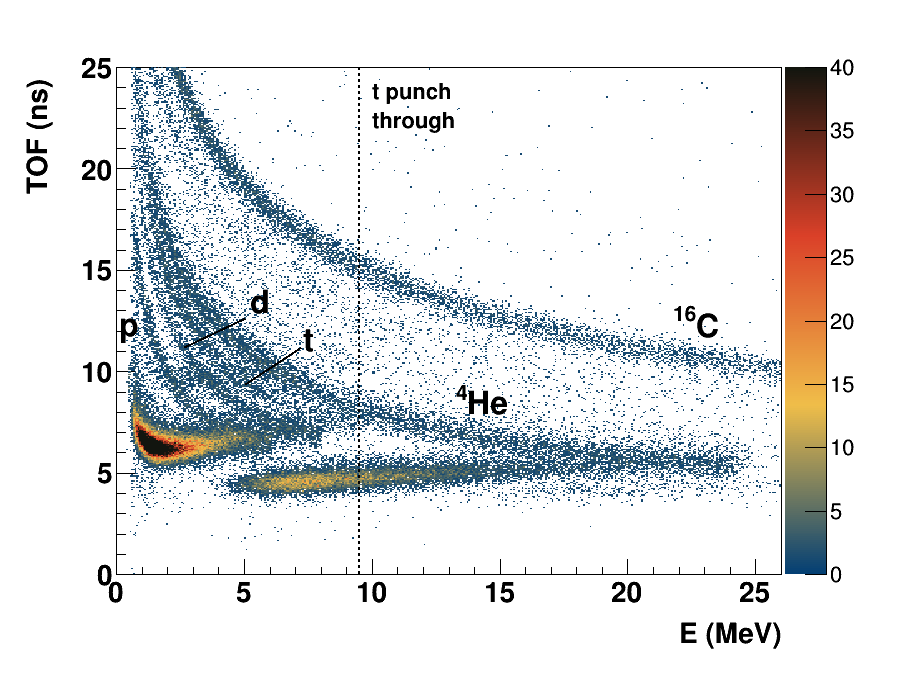}%
\caption{\label{E_TOF} Identification matrix: Time-Of-Flight (TOF) in ns between MUST2 and the first beam tracking detector versus the energy in MeV measured in the first layer of the MUST2 telescope. Dashed line shows the triton punch-through.}
\end{figure}
%\vspace{0.25mm}
\begin{figure}
  \vspace{0.15mm}
\includegraphics[scale=0.45]{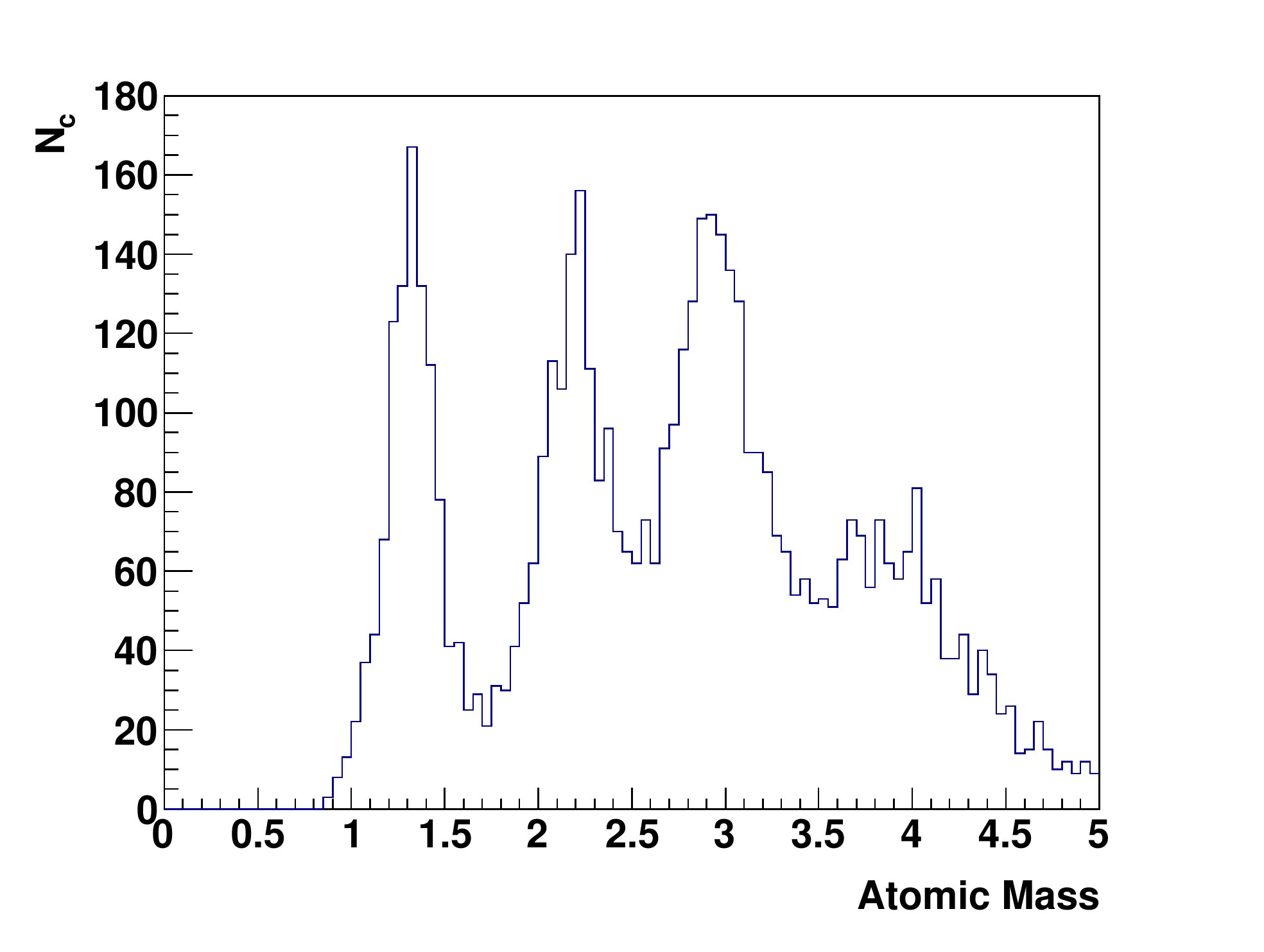}%
\caption{\label{mass_id} Experimental mass spectrum of the light ejectiles meausured in MUST2. The A=3 selection was taken in between 2.6 and 3.2}
\end{figure}

The experiment was performed at GANIL using a secondary $^{16}$C beam produced by fragmentation in the LISE spectrometer at 17.2~MeV/nucleon with an intensity of $\sim5 \times 10^{4}$ pps and  100\% purity. The target consisted of a CD$_{2}$ foil of 1.37(4) mg/cm$^2$. A set of two multiwire proportional chambers \cite{CATS} was used to track the beam onto the target on an event-by-event basis and served also as time reference and to measure the incoming beam intensity. The target was surrounded by the TIARA Silicon array \cite{TIARA} comprising an octagonal double-layered barrel of resistive strip detectors spanning 36$^{\circ}$ to 144$^{\circ}$ and an annular double-sided silicon strip detector covering the most backward angles (144$^{\circ}$ - 169$^{\circ}$). Light-ejectiles emitted from the neutron removal reaction $^{16}$C$(d,t)^{15}$C were detected by three MUST2 telescopes \cite{MUST2} placed in the forward hemisphere at 15 cm downstream the target covering angles from 10$^\circ$ to 40$^\circ$ in the laboratory frame. The  first layer of the MUST2 telescopes consisted of 300-$\mu$m thick double-sided stripped Silicon detectors (DSSD) that provided measurements of the total energy, the angle and served as the start signal for the TOF. The stop signal was obtained from the multiwire proportional chambers placed upstream the target. The beam-like fragments were detected in a Si-Si-CsI telescope located at zero degrees 33 cm downstream the target \cite{CHARISSA}. Only atomic-number identification was achieved during the experiment.  Gamma-rays from the dexcitation of the recoil were measured in coincidence in the four germanium clover detectors of the EXOGAM array \cite{EXOGAM} placed at 90$^\circ$ surrounding the barrel placed at a distance of 55 mm from the centre of the target. The only bound excited state in $^{15}$C is a state at 0.740 MeV with a half-life of 2.61 ns \cite{ENSDF} similar to the flight time from the target to MUST2. Additionally, the unbound states measured in this work are below the first excited state in $^{14}$C (6.09 MeV \cite{ENSDF}) and therefore no $\gamma$-rays were observed in coincidence with the tritons. Other results from this experimental campaign have been published previously, thus more details on the experimental set-up can be found in references \cite{XPereiraPLB,XPereiraPhD,BlCrom}.

Tritons from the $(d,t)$ reaction, which stopped in the DSSD, have been identified using the E-TOF technique.
Figure \ref{E_TOF} shows the E-TOF plot for the particles stopping in the DSSD of MUST2. Atomic mass information can be obtained from the following expression, knowing the flight path of each particle from the target to the telescope.
\begin{equation}
  A=\frac{2E}{u}\left( \frac{TOF}{L} \right)^{2}
  \label{eq_mass}
\end{equation}
where $E$ is the kinetic energy, $u$ the atomic mass unit, $L$ the flight path obtained from the positions of the particle in MUST2 and of the beam ion on target. Figure \ref{mass_id} shows the mass identification for particles with total energy E $<$ 8 MeV. Note that mass calibration was done for tritons and therefore the other peaks are slightly shifted. While only mass-separation is possible with this method, contamination from $^{3}$He particles is not expected below 8 MeV owing to the kinematics of the $^{16}$C$(d,^{3}He)$ reaction.

The dependence of the resolution with the excitation energy was determined using \textit{nptool} simulations package \cite{nptool}. The simulation included appart from the straggling of the particles in the different material layers and in the target, the geometry of the setup, the resolution of the different detectors, the secondary beam energy spread and spot size. 
%while the absolute value was obtained assuming the known value \cite{Goss2} of the natural decay width ($\Gamma_{n}$ = 75 keV) for the resonance at 3.10 MeV.

\begin{figure}
\includegraphics[scale=0.45]{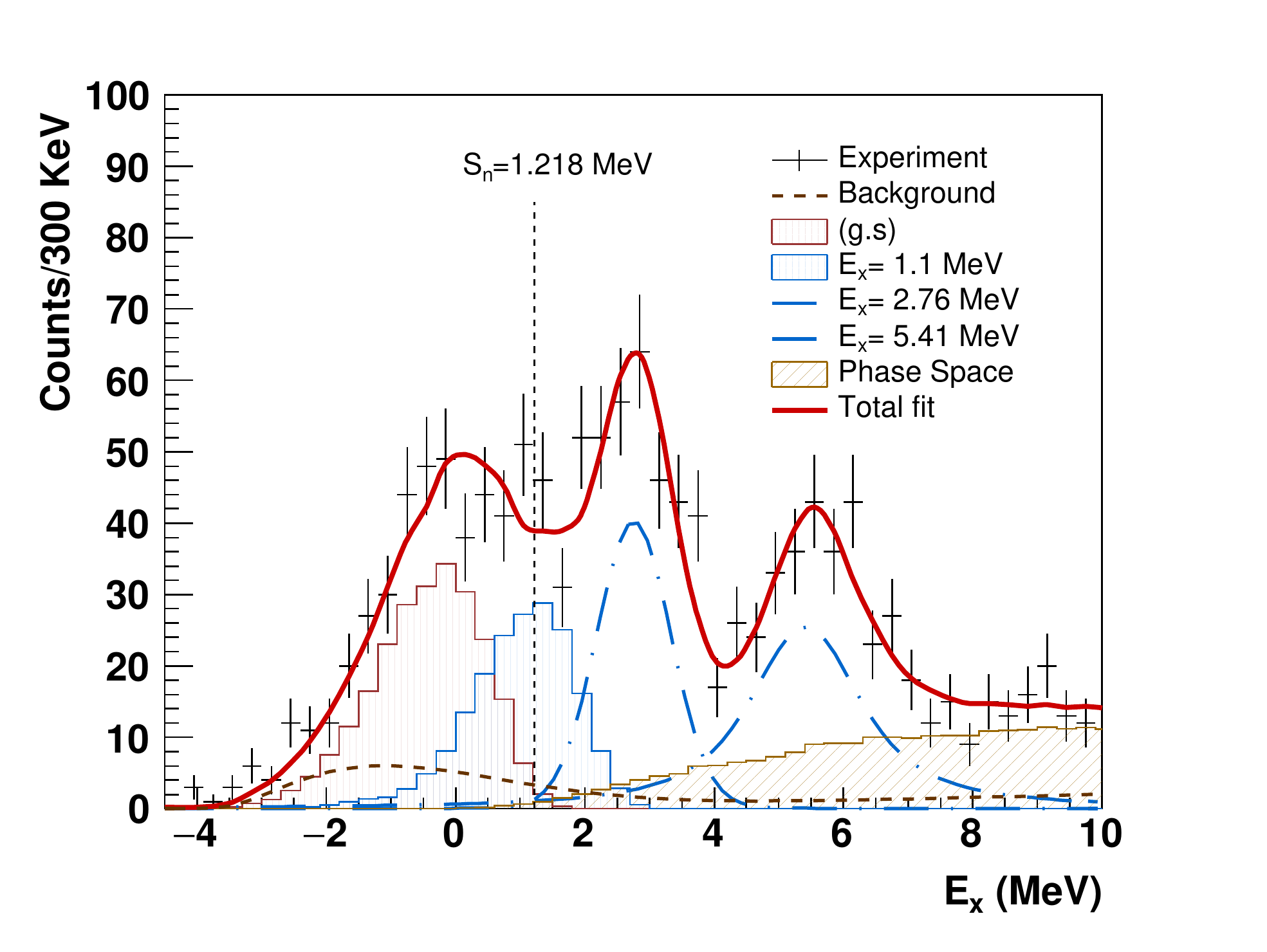}%
\caption{\label{exnrj} Experimental excitation energy spectrum for $^{15}$C obtained in the reaction $^{16}$C$(d,t)^{15}$C for the angular range 3$^{\circ}$-15$^{\circ}$ in the center of mass. Bound states are represented by profile histograms obtained from \textit{nptool} \cite{nptool} while dash-dotted curves correspond to the unbound states. The light brown filled histogram shows the contribution from the phase-space calculation and the brown dashed line the one from the background. The solid-red curve is the total fit to the spectrum. }
\end{figure}

\section{Results}

\begin{figure}
\includegraphics[scale=0.5]{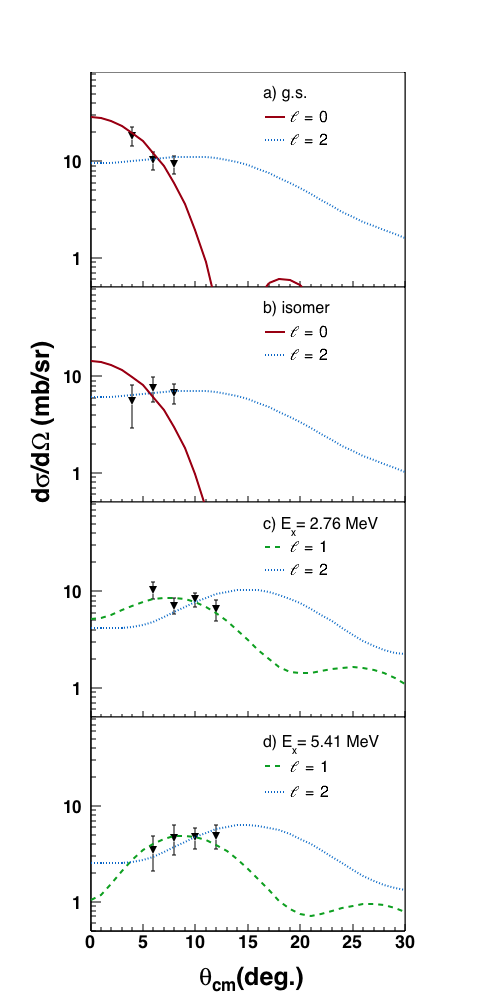}
\caption{\label{xs} Angular distributions for the ground state (a) the first excited state (b) and the two observed resonances in $^{15}$C at 2.76 (c) and 5.41 (d) MeV compared to $\ell$ = 0, 1, 2 (red, green, and blue) DWBA calculations. The uncertainties in the angular distributions are only statistical. The displayed cross-sections are not normalized to the occupancy of the $sd$-shell.}
\end{figure}

The $(d,t)$ events were selected as coincidences of a A=3 particle with energy below 8 MeV and a Z=6 heavy residue in the zero-degree telescope. Excitation energy from the single-neutron pickup channel was reconstructed using the missing-mass technique and the resulting spectrum in the center-of mass angular range between 3$^{\circ}$-15$^{\circ}$  is displayed in Fig. \ref{exnrj}. From the $(d,t)$ data, three clear peaks are observed in Figure \ref{exnrj}. Due to the low-energy of the tritons (Q$_{0}$=2.0 MeV) and their energy-loss straggling in the target, the resolution in excitation energy is not sufficient to separate the individual contributions below the neutron-separation threshold ($S_{n}$=1.218 MeV). In fact, the broad peak around 0 MeV corresponds to the 1/2$^{+}$ ground state and the 5/2$^{+}$ isomeric state at 0.74 MeV of $^{15}$C \cite{ENSDF}. Therefore, below the neutron-separation threshold the bound states were represented by two profile histograms that accounted for the experimental distortions of the shape. Above S$_{n}$, two Voigt functions were considered for the unbound states. Additionally, the contribution from a three-body final state (t+$^{14}$C+n) reaction has also been considered above the neutron-decay channel together with the background from masses with A=2 and 4 shown in Figure \ref{mass_id}.

From the $(d,t)$ data, four states are observed experimentally in $^{15}$C: the ground state, the isomeric state and two states at 2.76(7) and 5.41(11) MeV.
%which are assigned with the removal of a neutron from the 1$s_{1/2}$, 0$d_{5/2}$, 0$p_{1/2}$ and 0$p_{3/2}$ orbitals, respectively.
The resonance at 2.76 MeV was already observed in the $^{14}$C$(d,p)^{15}$C \cite{Goss2}, $^{13}$C$(t,p)^{15}$C \cite{Truong} and $^{13}$C$(^{18}$O$,^{16}$O$)^{15}$C \cite{Cappuzzello} experiments at approximately 3.10 MeV and was assigned to a spin and parity of 1/2$^{-}$. For the 5.41 MeV resonance, several states were detected around this region but not definitive assignment was made.
%(Check truong and goss2)

\begin{table*}
  \centering
  %\ra{1.3}
  \caption{\label{results} Spin-parity assignments $J^\pi$, excitation energies $E_{x}$ (keV), spectroscopic factors $C^2S$ normalized to the total occupancy of the $sd$-orbitals, asymptotic normalization coefficients squared (C$^{2}_{n}$) and corresponding transferred angular momenta $\ell$ for the states observed in $^{15}$C, compared to previous experimental data and shell model predictions with the SFO-tls interaction and to results of SCGF calculations using the NNLO$_{sat}$ interaction. The quoted uncertainties include statistical and fitting errors, systematic uncertainties from the choice of optical potentials (20 \%) and the error on the target thickness (3 \%). An additional error in the absolute value of the excitation energy, mainly due to the energy straggling in the target, was estimated to be 0.3 MeV.}
  \vspace{0.15cm}
  %\begin{tabular}{@{}p{1.1cm}p{1.2 cm}p{1.1 cm}p{1.3 cm}p{0.15cm}p{0.5 cm}p{1.5 cm}p{1.5 cm}p{1.2 cm}p{1.0 cm}p{1.2 cm}p{1.0 cm}@{}}
    \begin{tabular}{ccc ccc ccc cccc}
  \hline
  \hline
  \T\T
  %E$_{x}$(keV) & E$_{x}$(keV) & E$_{x}$(keV) & E$_{x}$(keV)  & $\ell$ & $J^{\pi}$ & $(2J+1)C^2S$ & $(2J+1)C^2S$ & E$_{x}$(keV) & $(2J+1)C^2S$ & E$_{x}$(keV) & $(2J+1)C^2S$ \\
  E$_{x}$ (keV) & E$_{x}$ (keV) & E$_{x}$ (keV)  & $\ell$ & $J^{\pi}$ & $C^2S$ & C$^{2}_{n}$ (fm$^{-1}$) &$C^2S$ & $J^{\pi}$ & E$_{x}$ (keV) & $C^2S$ & E$_{x}$ (keV) & $C^2S$ \\
  \cmidrule{3-7}
  \cmidrule{9-10}
  \cmidrule{11-13}
  $(d,p)$     & Knockout   & \multicolumn{5}{c}{$(d,t)$} & Knockout  & &\multicolumn{2}{c}{SFO-tls} &  \multicolumn{2}{c}{NNLSO$_{sat}$} \\
   \cite{Goss2}& \cite{Maddalena} &   \multicolumn{5}{c}{(present)} & \cite{Maddalena} &    &   &   & &\T\B\\
  
  \hline
  0 & 0 & -0.30$_{-(22)}^{+(18)}$ & 0 & $1/2^{+}$ & 0.65(16) & 8.66(17) & 0.56(10)   & $1/2^{+}$ & 0.0 & 0.64  & 0.168 & 0.51  \T\\
  0.744 & 0.740 & 1.10$_{-(16)}^{+(28)}$ & 2 & $5/2^{+}$ & 1.35(32) & 1.56(3) & 1.28(20) & $5/2^{+}$ & 0.854 & 1.20 & 0.00 &  1.10\\
  3.105 &      & 2.76(7) & 1 & $1/2^{-}$ & 1.74(23) & 15.78(31) &   & $1/2^{-}$  & 2.74 & 1.35  & 6.62 & 0.35 \\
  &           &         &   &    &  &   & & $1/2^{-}$  &  4.71 & 0.10  & 7.88  & 1.11  \\
  &            & 5.41(11) & 1 & $3/2^{-}$ & 1.25(24) & 26.85(54) &  & $3/2^{-}$   & 5.30 & 0.62 & 10.63  & 0.63\\
  &              &           &   &          &           &  & & $3/2^{-}$   & 6.00 & 0.68 & 10.80  & 1.14 \B\\
 % &       &       &         & 1 & $3/2^{-}$ & 0.86(23) &     & 5.296 & 0.62 & --    & -- \\
%  &       &       &         & 1 & $3/2^{-}$ & 0.86(23) &     & 5.296 & 0.62 & --    & -- \B\\

		\hline
                \hline
\end{tabular}
\end{table*}

Differential cross-sections for 3-4 angular-bins of 2$^{\circ}$ width each have been reconstructed from the integral of each line-shape. The angular range originnally from 3$^{\circ}$-15$^{\circ}$ in the center of mass is limited in order to avoid the region of abrupt changes in acceptance. The angular distributions from the $(d,t)$ reaction are shown in Figure \ref{xs}, and are compared to finite-range Distorted-Wave Born Approximation (DWBA) calculations using the code FRESCO \cite{FRESCO} to deduce the spectroscopic factors. 
%For the $d$+$^{16}$C entrance channel, the optical model parameters used in the analysis were obtained from the modified global parameterisation of Haixia et al.\cite{Haixia} shown in \cite{XPereiraPhD}.
For the $d$+$^{16}$C entrance channel, the optical model parameters used in the analysis were obtained from the global parameterisation of Haixia et al. \cite{Haixia}, modified to reproduce $d$+$^{16}$C elastic scattering measured along with $^{16}$C$(d,t)$ \cite{XPereiraPhD}.
These elastic data showed that an increased diffuseness of the imaginary part of the optical potential was needed \cite{XPereiraPhD}. For the $t$+$^{15}$C exit channel the optical model potential from Pang et al. \cite{Pang}, was employed.
The $\langle d  |  t \rangle$ overlap was computed in a potential reproducing results from Green's function Monte Carlo calculations \cite{Brida}.
%Ab-initio calculations developped by I. Brida et al. \cite{Brida}, using a Quantum Monte Carlo method (GFMC) and realistic two-neutron and three-neutron potentials were used for the $<d ~\vrule~ t>$ vertex. %Standard values for the vertex $<d ~\vrule~ t>$ constant D$_{0}^{2}$=25600 MeV$^{2}$ fm$^{3}$ and the Hulten finite range of 0.746269 fm were chosen.
%(Prior form and no remannant)
The $\langle {^{16}\mbox{C}} | ^{15}\mbox{C} \rangle$ overlaps were represented by bound neutron wave functions obtained in a Woods-Saxon potential with standard geometry (radius parameter $r_{0}$ = 1.25 fm and diffuseness $a$ = 0.65 fm), and with the depth adjusted to reproduce the effective neutron separation energy.
%The bound state wave function was obtained with a standard Woods-Saxon potential where the depth of the potential was adjusted to the effective excitation energy of each state and with a reduced radius r$_{0}$ = 1.25 fm, and diffuseness a$_{0}$ = 0.65 fm.
%ifferent combinations of deuteron-nucleus and triton-nucleus potentials were used to estimate the uncertainty introduced by the choice of potentials arising to $\pm$20\%.
%\textit{Different combinations of deuteron-nucleus and triton-nucleus potentials were used to estimate the uncertainty introduced by the choice of potentials arising to 20\%. This is not absolutely true....}

From the shape of the angular distributions the orbital angular momentum of the removed nucleon is deduced. Relative spectroscopic factors were obtained by normalizing the theoretical curves to the experimental data. Absolute information on the occupancy of each single-particle orbital was determined by requiring the sum of the 1/2$^{+}$ and 5/2$^{+}$ states to be equal to 2, which is the expected occupancy of the neutrons in the $sd$-shell. This procedure disregards the occupancy of the 0$d_{3/2}$ orbital. However, this contribution is expected to be negligeable. In fact in the near isotope $^{17}$C \cite{XPereiraPLB}, the 3/2$^{+}$ ground state is interpreted as three neutrons in the 0$d_{5/2}$ orbital and no single-particle 3/2$^{+}$ state has been located so far. The results are displayed in Table \ref{results}. The uncertainties in the determination of the spectroscopic factors arise from statistical and fitting errors, and the uncertainty in the target thickness (3 \%). Additionally, different combinations of deuteron-nucleus and triton-nucleus potentials were used to estimate the uncertainty introduced by the choice of potentials arising to $\pm$20\%. Uncertainties in the excitation energy due to the energy straggling in the target was estimated to be 0.3 MeV.

%In addition, inspired by \cite{Akra}, we have extracted the asymtotic normalization coefficient from the $^{16}$C knockout reaction \cite{Maddalena,Yamaguchi} to the ground and first excited state in order to cross check the normalisation. 

Top panel of Fig. \ref{xs} shows a $\ell=0$ character of the angular distribution for the ground state $[$see Fig. \ref{xs} (a)$]$  which is consistent with the known assignment of 1/2$^{+}$ made in \cite{Maddalena,Yamaguchi,Kay}. The angular distribution for the peak at approximately 0.74 MeV $[$see Fig. \ref{xs} (b)$]$  displays an $\ell=2$ contribution in agreement with the 5/2$^{+}$ assignment by Refs. \cite{Maddalena,Yamaguchi,Kay}.
%The relative strength of the $($1$s_{1/2})^{2}$ and $($0$d_{5/2})^{2}$ configurations obtained for $^{16}$C is very similar to the one obtained in the $^{18}$O$(d,t)$ reaction where the spectroscopic factors are $C^{2}S=$1.53 for the 0$d_{5/2}$ and $C^{2}S=$0.21 for the 1$s_{1/2}$ orbital.

%The summed number of $sd$ neutrons in the ground state of $^{16}$C amounts to 1.35 (--) which is lower than the expected value of 2 owing to the deformation of the $^{16}$C ground state \cite{Hamamoto}.
%As mentioned above, the experimental resolution prevent us from separating the ground and isomeric state and therefore only the angular distribution for the sum of the bound states is shown in top pannel of Figure \ref{xs}. Since the spin and parity of the ground and isomeric states are well known, the spectroscopic factors are deduced by a linear combination of the theoretical DWBA calculations for $\ell$=0 and $\ell$=2. The obtained values are 0.25 and 1.46, respectively.  The experimental angular distribution is well described by ...or displays l and. 

Figure \ref{xs} (c) shows the angular distribution for the first resonance at 2.76(7) MeV. The $\ell=2$ curve yields an unphysical occupancy of the state.  The available strength of the $0d_{5/2}$ orbital for the single-neutron pick-up is almost exhausted in the isomeric state at 0.74 MeV as the remaining has been observed in the $^{16}$C$(d,p)^{17}$C reaction \cite{XPereiraPLB}. Based on the Macfarlane-French sum rule for C$^{2}$S \cite{MF-sumrule}, where the occupancy is $G^{-}(0d_{5/2})$=1.35(32) (see Table \ref{results}) and the vancancy is $G^{+}(0d_{5/2})$=$\frac{2J_{f}+1}{2J_{i}+1}$=6$\times$0.62(13) (\cite{XPereiraPLB}), the maximum single-particle strength accounts up to 84.5(88)\% of the total. In addition, the $\ell=1$ transfer is favoured by the data. Therefore, the spin and parity of the 2.76(7) MeV is set to be 1/2$^{-}$ in agreement with earlier studies \cite{Goss2, Truong, Cappuzzello}.
%As shown by the deduced spectroscopic factor (see Table \ref{xs}), the 1/2$^{-}$ state contains a major part of the total 0$p_{1/2}$ strength.
The configuration of this state will thus correspond to a neutron removal from the $0p_{1/2}$ orbital in $^{16}$C.
%We note that in the N=9 isotone $^{17}$O, the first 1/2$^{-}$ state was observed at 3.055 MeV with a similar spectroscopic strength of $C^{2}S=$1.08 \cite{Mairle}

The angular distribution presented in the bottom panel  $[$see Fig. \ref{xs} (d)$]$ corresponds to the state at 5.41(11) MeV. Only the comparison to the  $\ell=1$ calculation yields a reasonable spectroscopic factor, while the value for $\ell=2$ is incompatible with an acceptable occupation of the 0$d_{5/2}$ or 0$d_{3/2}$ orbitals in the ground state wave function of $^{16}$C. Besides, assuming another 1/2$^{-}$ assignment for this state, the summed C$^{2}$S would largely exceed the maximum number of neutrons in the $0p_{1/2}$ orbital. Given the large spectroscopic factor and guided by the structure calculations presented in Table \ref{results}, we assign this state at 5.41 MeV as a neutron 0$p_{3/2}$ hole and therefore a spin and parity of 3/2$^{-}$. Previous results in references \cite{Truong,Cappuzzello} and \cite{Miyatake} point to an state at 5.88 MeV with a tentative 1/2$^{-}$ assignment but 3/2$^{-}$ could not be totally excluded. Owing to the high excitation energy of the first excited state in $^{14}$C, the resonance states populated in $^{15}$C in the $(d,t)$ reaction can only decay by neutron emission to the ground state of $^{14}$C.
%Because of the limited experimental resolution and uncertainty in the amount of phase-space contribution to the spectrum, only estimates for the $\Gamma_{n}$ widths are given in the present work.

\begin{figure}
\includegraphics[scale=0.45]{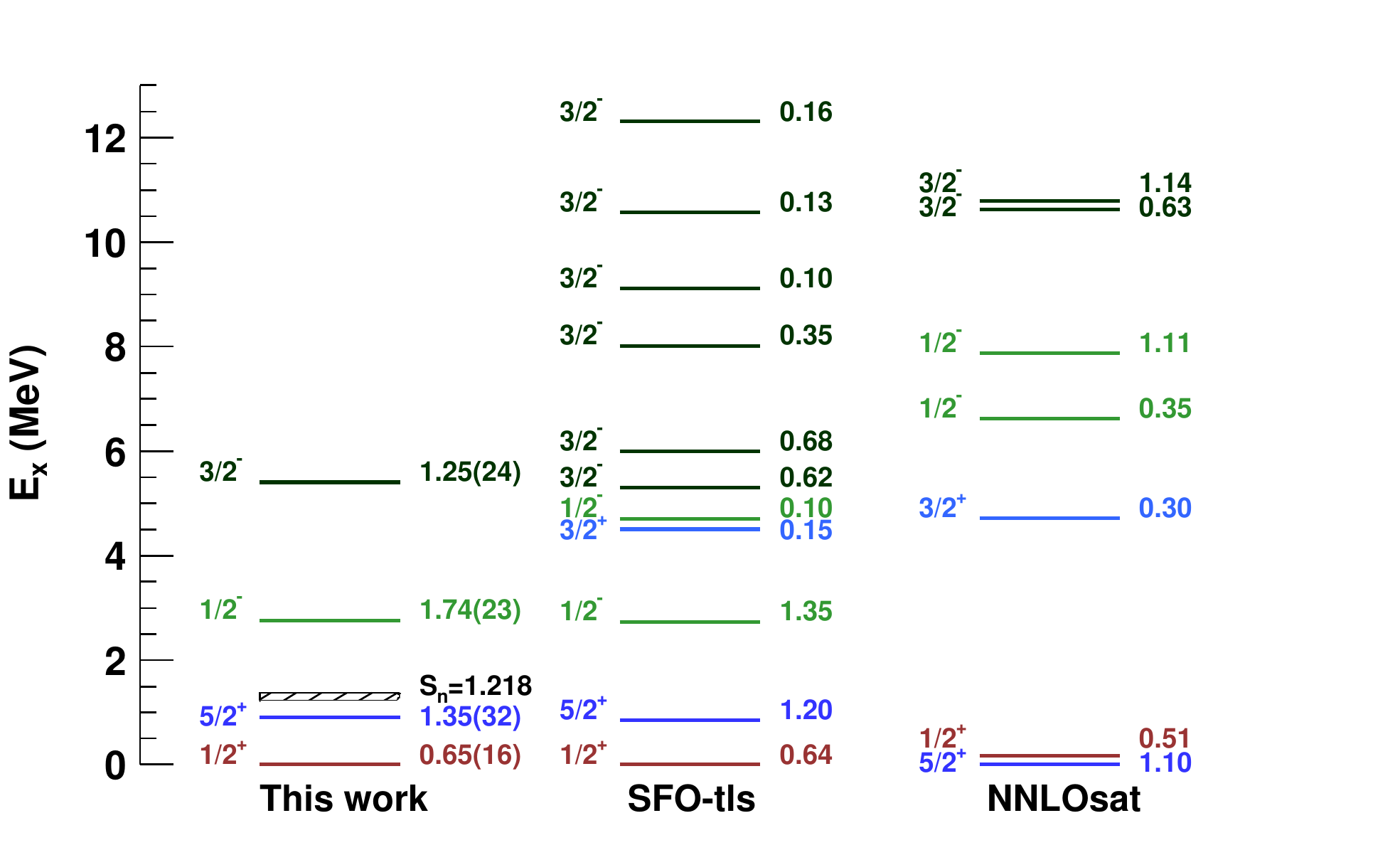}%
\caption{\label{nrj_comp} Experimental level scheme for $^{15}$C compared to the results of the shell model with the SFO-tls interaction and to SCGF calculations using the NNLO$_{sat}$ interaction.
  %The dashed line links states with similar spectroscopic strength. Only those states with spectroscopic factors above 0.10 have been displayed in the Figure.
  The column on the right represents the spectroscopic factor and on the left the spin and parity assignment. }
\end{figure}

Pick-up reactions such as $(d,t)$ are of peripheral nature and represent a suitable tool to extract information on the asymptotic tail of the radial overlap function which is very sensitive to the $NN$ interactions used in microscopic calculations \cite{BAYE}. The asymptotic normalization coefficient (ANC) squared C$^{2}_{n}$ was obtained as the product of the experimental spectroscopic factor S$_{lj}$ and the square of the single-particle ANC b$_{lj}$. Spectroscopic factors extracted from peripheral reactions are usually sensitive to the choice of the neutron-nucleus potential used to generate the bound state wave function of the transferred nucleon \cite{BFD21Al,BFD25P}. In order to estimate this effect, the radius of the single-particle Woods-Saxon potential was varied from 1.15 to 1.35 fm and the resulting spectroscopic factors changed with respect to the standard value of 1.25 fm as follows: 35\% (0$d_{5/2}$), 10\% (1$s_{1/2}$), 20\% (0$p_{1/2}$), 25\% (0$p_{3/2}$) while the ANC remained constant within 1-3\%, showing the peripheral character of the reaction.  Therefore, in addition to the spectroscopic factors, the ANCs for the four states populated in $^{15}$C are also reported in this work (see Table \ref{results}).

\section{Discussion}

The results of the $(d,t)$ reaction are displayed in Table \ref{results} and in Fig. \ref{nrj_comp}. The experimental level scheme is compared to state-of-the art shell-model calculations using the SFO-tls interaction \cite{TSuzukiI,TSuzukiII} within the $p$-$sd$ model space. This interaction has proven to be successful in reproducing the level ordering and single-neutron spectroscopic factors \cite{XPereiraPLB} as well as the magnetic dipole transitions in $^{17}$C \cite{TSuzukiII}. Additionally, we compare our results with \textit{ab initio} many-body calculations using the self-consistent Green\textquotesingle s function (SCGF) method at second order \cite{VSoma1,VSoma2} with the NNLO$_{sat}$ interaction \cite{Ekstrom} from chiral effective field theory. The two-nucleon and three-nucleon forces were optimised to reproduce low-energy experimental observables in order to provide improved predictions for mid-mass nuclei \cite{Ekstrom}. In particular, the $NN$+$3N$ fit at next-to-next-to-leading order includes, apart from the nucleon-nucleon scattering data, the binding energies of $^{3}$H, $^{3,4}$He,$^{14}$C, and $^{16,22,24,25}$O together with the charge radii of the first five isotopes. The resulting interaction NNLO$_{sat}$ improves significantly the properties of mid-mass nuclei and reproduces the saturation point of nuclear matter \cite{Ekstrom,Lapoux,VSoma3,VSoma4} but has seldomly been tested for the spectroscopy of $p-sd$ shell nuclei.

\begin{figure*}[h]
\centering
\begin{minipage}{.45\linewidth}
  \includegraphics[width=\linewidth]{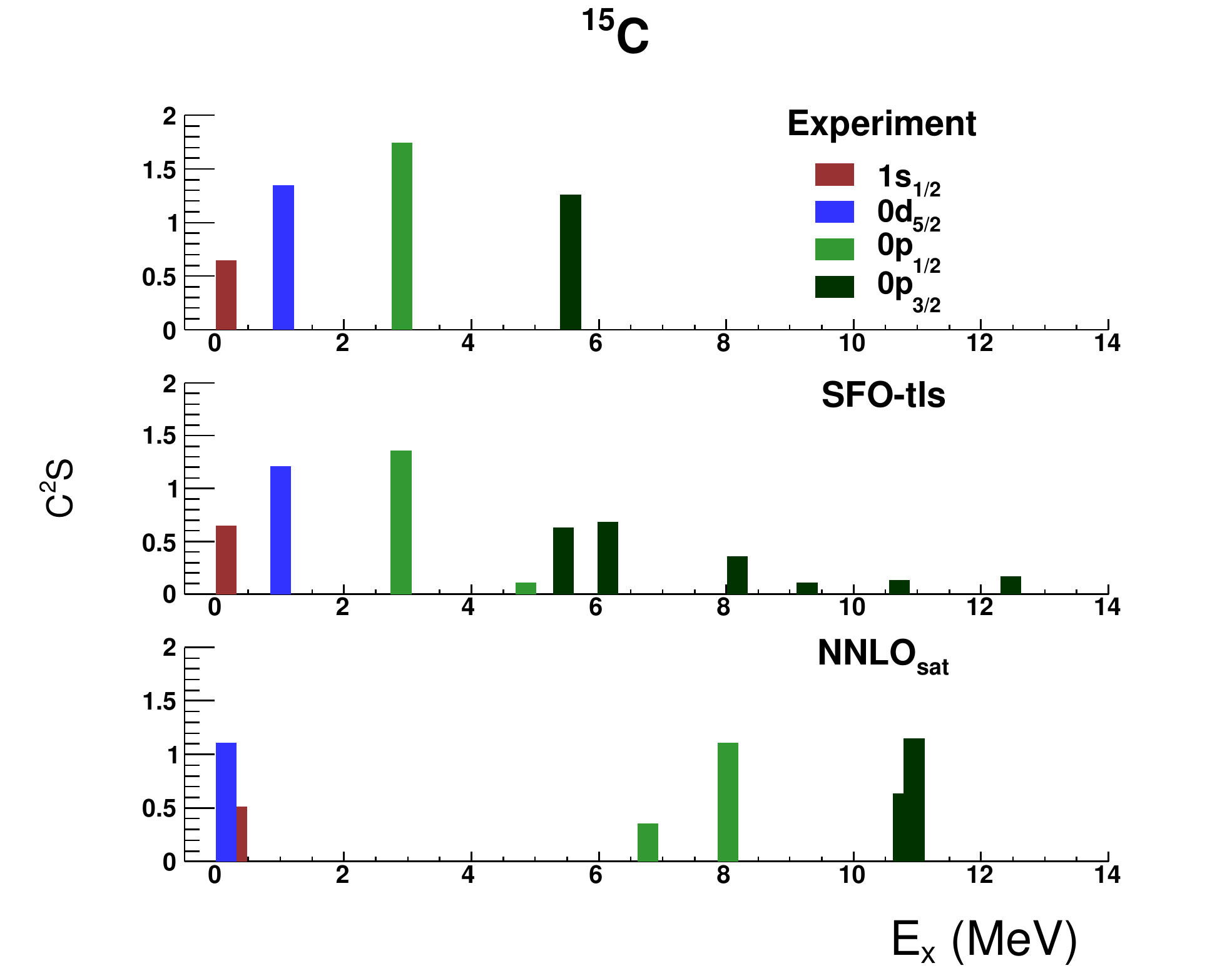}
%  \caption{$^{15}$C}
  %\label{centroid_15C}
 \end{minipage}
\hspace{.05\linewidth}
\begin{minipage}{.45\linewidth}
    \includegraphics[width=\linewidth]{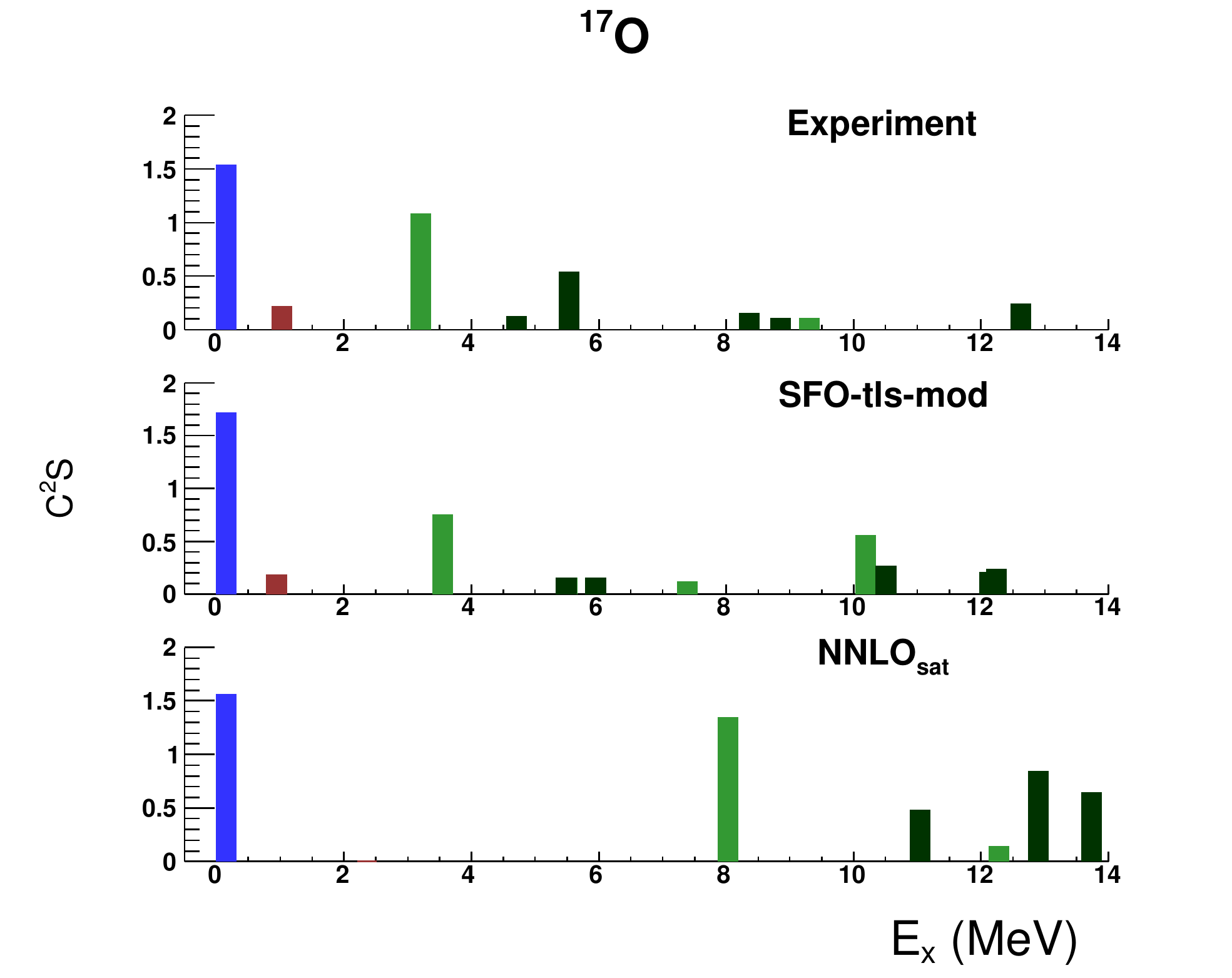}
%    \caption{$^{17}$O}
    %\label{centroid_17O}
\end{minipage}
\caption{Distribution of the measured single-particle strength $C^2S$ for $^{15}$C (left) and $^{17}$O (right) as a function of excitation energy in MeV. Data from the $^{18}$O$(d,t)^{17}$O reaction correspond to reference \cite{Mairle}. }
  %The position of the arrows shows the average strength}

\label{centroid_N9}
\end{figure*}

%an advance parameterisation of a Hamiltonian based on chiral effective field theory labeled NNLO$_{sat}$, that provides improved predictions of nuclear saturation and a better reproduction of charge radii data \cite{VSoma}. 

Positive-parity states in $^{15}$C measure the 1p configurations. For the ground state and first excited state, the obtained spectroscopic factors are in good agreement with the results from \cite{Maddalena,Yamaguchi} within the error bars  $[$see Fig. \ref{nrj_comp}$]$. Both values are well reproduced by the shell model with the SFO-tls interaction taking into account the experimental uncertainties. As for the SCGF calculations employing the NNLO$_{sat}$ interaction, the 1/2$^{+}$ and 5/2${+}$ states are almost degenerate in energy. However, the spectroscopic factors for the SCGF+NNLO$_{sat}$ interaction of the ground and isomeric state are in line with the experimental results.

%Since the ground and first excited states have been identified as single-particle 1$s_{1/2}$ and 0$d_{5/2}$ neutron levels outside a $^{14}$C core,
The deduced spectroscopic factors can be compared to the ones obtained in the $^{18}$O$(d,t)$ reaction by G. Mairle et al., \cite{Mairle} where the spectroscopic factors for the $C^{2}S$(0$d_{5/2}$) and $C^{2}S$(1$s_{1/2}$) are 1.53 and 0.21, respectively.  The raise of the 0$d_{5/2}$ spectroscopic strength in $^{17}$O, as we add protons to the 0$p_{1/2}$ orbital, shows an increase in the filling of the 0$d_{5/2}$ orbital, due to an inversion of the 0$d_{5/2}$ and 1$s_{1/2}$ effective single-particle energies as described in \cite{OtsukaRMP}. Nevertheless, the filling of the 0$d_{5/2}$ orbital in $^{16}$C remains dominant due to pairing effects. 

%can be used to obtain the relative strength of the $($0$d_{5/2})^{2}$ and $($1$s_{1/2})^{2}$ configurations in the ground state wave function of $^{16}$C. Assuming no significant admixture from other configurations \footnote{Here we have neglected the 0d$_{3/2}$ orbital and any other collective configurations}, the amplitude of the 0$d_{5/2}$ and 1$s_{1/2}$ components are 0.82 and 0.57, respectively. The obtained values can be compared to the ones obtained in the $^{18}$O$(d,t)$ reaction by G. Mairle et al., \cite{Mairle} where the spectroscopic factors for the $C^{2}S$(0$d_{5/2}$)=1.53 and $C^{2}S$(1$s_{1/2}$)=0.21, yielded amplitudes of 0.87 and 0.32 for each of the configurations.  A later experiment of the same reaction $^{18}$O$(d,t)$ by H.T. Fortune et al., \cite{Fortunesd} using a DWBA analysis concluded with comparable results. The stronger 1$(s_{1/2})^{2}$ configuration in $^{15}$C might be related to the near degeneracy of the 0$d_{5/2}$ and 1$s_{1/2}$  orbitals in neutron-rich carbon isotopes \cite{XPereiraPLB,MStanoiu-N=14Cchain}
%The summed number of $sd$ neutrons in the ground state of $^{16}$C amounts to 1.35 (--) which is lower than the expected value of 2 probably due to the deformation of the $^{16}$C ground state \cite{Hamamoto}. In $^{17}$O, the sum is somewhat higher 1.74 as a result of a stronger 1$d_{5/2}$ component.

%a collapse of the N=14 shell gap \cite{MStanoiu-N=14Cchain, XPereiraPLB}
%Except for the 0.4 MeV difference in excitation energy, the state at 3.10 MeV is in very good agreement with the predicted energies using the SFO-tls interaction.

Above the neutron separation threshold, negative-parity states appear with a 2p-1h structure. As shown by the experimental spectroscopic factor (see Table \ref{results}), the 1/2$^{-}$ state contains a significant part of the total 0$p_{1/2}$ strength. In fact, the observed strength for this state represents 87\% of the total occupancy assuming an independent particle filling. The configuration of this state is interpreted to be a $^{13}$C (g.s.)$\times($1$s_{1/2}$0$d_{5/2})^{2}$ configuration in view of the large spectroscopic factor for the $(\nu 0p{_{1/2}})^{-1}$. The excitation energy of the 1/2$^{-}$ state provides some qualitative information on the size of the N=8 gap. While the shell model with the SFO-tls interaction reproduces the energy of the state that carries the main fragment of the $\nu0p_{1/2}$ orbital, the SCGF method with NNLO$_{sat}$ places this state at much higher energy $[$see Fig. \ref{nrj_comp}$]$. This is mainly due to the lack of 2p-1h configurations in the latter theoretical scheme, presently truncated at second order. Additionally, the NNLO$_{sat}$ interaction generates a large mean-field gap between the $p$ and $sd$ harmonic-oscillator shells, which further hinders the onset of the relevant cross-shell correlations.

Turning to the 0$p_{3/2}$ strength, as shown in Fig. \ref{nrj_comp} the theoretical strength distributions are significantly fragmented, especially in the shell-model calculations with SFO-tls. The experimental strength for the 0$p_{3/2}$ orbital is 1.25(24) compared to the independent single-particle expectation of 4, therefore indicative of a significant fragmentation, with a large fraction of the strength possibly at higher excitation energy. As shown in Fig. \ref{nrj_comp} there are several states predicted by shell model calculations above 8 MeV that could not be seen in our experiment due to the limited statistics and resolution. The experimental total strength thus represents a lower limit. The excitation energy of the strongest 3/2$^{-}$ state is in good agreement with the results from the shell model with the SFO-tls interaction, showing that the state we see might be a combination of the two resonances predicted in the region. Similarly, for the 1/2$^{-}$ state, the predictions from the SCGF calculations with NNLO$_{sat}$ appear at much higher excitation energy. Surprisingly, the total strength is in reasonable agreement with the experimental results. It is important to note that the difference in energy between the major fragments of the 3/2$^{-}$ and 1/2$^{-}$ states is instead well reproduced  ($\Delta_{N=6} \approx$ 2.9 MeV) which suggests that the mechanism at play, and the corresponding missing correlations in SCGF calculations, are indeed the same for the two states.

The 2p-1h configurations of the 2.76 and 5.41 MeV states are clearly established. As shown by the deduced spectroscopic factors (see Table \ref{results}),  the 1/2$^{-}$ and  3/2$^{-}$ states contain a sizeable part of the total 0$p_{1/2}$ and 0$p_{3/2}$ strength (sd)$^{2}$(1p)$^{-1}$. Although part of the strength is expected at higher excitation energies we can compare our results with the ones obtained in the neutron removal from $^{18}$O in the same excitation energy regime.

Figure \ref{centroid_N9} shows the measured strength from the $(d,t)$ reaction as a function of the excitation energy for $^{15}$C and $^{17}$O. In the case of $^{15}$C, there is a very good agreement between the experimental distribution of the single particle strength and major fragments from the shell model using SFO-tls, with the exception of a higher degree of fragmentation in the shell-model 0$p_{3/2}$ strength, not observed in our experiment possibly because of limited statistics and resolution. While SCGF calculations using the NNLO$_{sat}$ Hamiltonian perform well with natural parity states, cross-shell states are however overpredicted in energy.

Very interesting is the comparison with the N=9 isotone $^{17}$O. Experimentally, an inversion of the 1/2$^{+}$ and 5/2$^{+}$ states is observed which points to a major change of structure between the 1$s_{1/2}$ and the 0$d_{5/2}$ as described in \cite{OtsukaRMP}.
Owing to the single-particle character, the excitation energy of the 1/2$^{-}$ state provides a measure of the N=8 shell gap. Experimentally, this state is observed at rather similar energies in both isotones 2.76(7) MeV in $^{15}$C and 3.055 MeV in $^{17}$O pointing to a comparable N=8 shell gap. However, the single-particle strength of the 1/2$^{-}$ state is more fragmented in $^{17}$O as a consequence of an increase of correlations in the core of the N=Z nucleus. Theoretically, shell model calculations using the SFO-tls interaction had to be modified in order to reproduce the $^{17}$O spectrum shown in Figure \ref{centroid_N9}. As a matter of fact, valence particles are more weakly bound in neutron-rich carbon isotopes compared with the oxygen isotopes. Note that the valence neutron in $^{15}$C has a halo structure with one-neutron separation energy of 1.218 MeV. Effective nuclear forces that involve weakly bound neutrons are expected to be weaker than those for normal nuclei \cite{MStanoiu-N=14Cchain}. As the SFO-tls is constructed, in particular, to explain the structure of neutron-rich carbon isotopes, it is a reasonable prescription to use stronger nuclear forces in oxygen isotopes. For the oxygen case, monopole terms of two-body matrix elements that involve valence particles in the $sd$-shell are enhanced compared to the carbon case. Distribution of the single-particle strength obtained is shown in the right panel of Figure \ref{centroid_N9}, where the monopole terms of $p-sd$ and $sd-sd$ matrix elements are more attractive by 0.375 MeV in the isospin T=0 channel and by 0.125 MeV in the isospin T=1 channels, respectively, than the original SFO-tls. The calculated distribution of the strength with the modified SFO-tls interaction (SFO-tls-mod) is found to be consistent with the experimental data. The values obtained for the N=8 shell gap in the shell model calculations are 9.05 MeV in $^{15}$C (with SFO-tls) and 8.33 MeV in $^{17}$O (with SFO-tls-mod) pointing to a rather constant value as observed experimentally. Concerning the 3/2$^{-}$ state, the major fragment of the strength appears at excitation energies around $~$ 5.5 MeV in the experimental spectra of $^{15}$C and $^{17}$O. The distribution of the strength for the 3/2$^{-}$ states predicted theoretically by shell model calculations using the SFO-tls and SFO-tls-mod interactions seem to split the occupancy in two states at rather close energies. As for the N=6 gap, it becomes difficult to draw any conclusions considering that only about 1/3 of the total strength has been observed in our experiment and it appears to be fragmented in several states. 

%In this case, the SFO-tls interaction which gave a good account of the $^{15}$C 2p-1h structure fails in reproducing the size of the N=8 shell gap and also the position of the centroid for the 0$p_{3/2}$ orbital appears to be shifted at higher energy. The large difference in energy for the negative-parity states in $^{17}$O might be explained as a consequence of their internal configuration. The structure of the 1/2$^{-}$ and 3/2$^{-}$ states can be interpreted as $^{16}$O(3$^{-}$) $\times$ 0$d_{5/2}$ in the weak coupling model. In fact, experimentally the excitation energy of the first 3$^{-}$ state in $^{16}$O is found to 6.13 MeV \cite{ENSDF} while the SFO-tls interaction predicts it to be at 7.633 MeV, which is 1.5 MeV higher than expected. An improved description in the energy of this state may lead to a much more accurate description of the N and O isotopes as stated in C. Yuan et al., \cite{YSOX}.  
Regarding the results of $^{17}$O obtained from SCGF calculations with NNLO$_{sat}$, we observe a similar behaviour as in $^{15}$C, \textit{i.e.} an overestimation of the N=8 whereas N=6 is fairly well reproduced, suggesting a general trend from Z=6 to Z=8. Moreover, when going from $^{17}$O to $^{15}$C, the increase in fragmentation of the 5/2$^{+}$ state and the associated population of the 1/2$^{+}$ state are also well described. Conversely, the phenomenological SFO-tls interaction works locally in the region of Z=6 but further modifications in the monopole terms are needed when applying the same Hamiltonian to other  $p-sd$ regions of the nuclear chart. 

\section{Summary}
In summary, the results of the $^{16}$C$(d,t)^{15}$C reaction presented here bring new information on the $p$-$sd$ orbitals in $^{15}$C. Positive parity states with a 1p configuration in $^{15}$C confirm the inversion of the 1$s_{1/2}$ and 0$d_{5/2}$ orbitals. The variation in the relative strength of the single-particle orbits 1$s_{1/2}$ and 0$d_{5/2}$ along the N=9 isotonic chain, from $^{15}$C to $^{17}$O, is a combined effect of the shell evolution mechanism due to the monopole interaction explained in \cite{OtsukaRMP} and the pairing contribution.   

The relative position of the 1/2$^{-}$ and 3/2$^{-}$ levels, with respect to the positive-parity states in the $sd$-shell, is sensitive to the size of the N=8 shell gap. Calculations using the SFO-tls interaction agree well with the results obtained here for $^{15}$C. However, the comparison with its isotone $^{17}$O at N=9, showed the need of modifying the monopole terms of the $p$-$sd$ and $sd$-$sd$ two-body matrix elements by 0.375 MeV in the T=0 channel and by 0.125 MeV in the T=1 channel in order to get an improved description in the $p$-$sd$ region of the nuclear chart.

While \textit{ab initio} SCGF calculations reproduce the fragmentation of positive-parity states in $^{15}$C, negative-parity states appear too high in energy. This highlights an enlarge N=8 shell gap and missing 2p-1h correlations in the current level of approximation and provides a strong motivation for extending the approach to higher orders \cite{barbieri22}.

%\begin{acknowledgments}
\section*{Acknowledgments}
J.L.F wishes to acknowledge financial support from Xunta de Galicia (Spain) grant number ED481A-2020/069, X.P.L. acknowledges support by IN2P3/CNRS (France) doctoral fellowship, the ST/P003885 grant (Spain) and Grant No. IBS-R031-D1. This work was supported by the Spanish MINECO through the project PGC2018-096717-B-C22 and PID2021-128487NB-I00. This work was also partially supported by the Xunta de Galicia under project No ED431B 2018/015, 2021-PG045 and the Maria de Maeztu Unit of Excellence MDM-2016-0692. W.N.C., N.T. and A.M. acknowledge financial support from the STFC grant number ST/L005743/1. M.F., N.C., Tz.K. and C.W acknowledge support from STFC grant number ST/V001043/1. The authors acknowledge the support provided by the technical staff of LPC-Caen and GANIL. SCGF calculations were performed by using HPC resources from GENCI-TGCC (Contract no. A0110513012). The participants from the Universities of Birmingham and Surrey, as well as the INFN and IFIN-HH laboratories also acknowledge partial support from the European Community within the FP6 contract EURONS RII3-CT-2004-06065.

%\end{acknowledgments}

%\bibliographystyle{elsarticle-num}
%\bibliography{biblio_v2}

\begin{thebibliography}{99}

\bibitem{OtsukaRMP}
  {T. Otsuka, A. Gade, O. Sorlin, T. Suzuki and Y. Utsuno},
  %title = {Nuclear magic numbers: New features far from stability},
  {Rev. Mod. Phys.}
   {92} 
   {(2020)}
   {015002}.
   
\bibitem{Bazin}
  {D. Bazin et al.},
  %title = {Probing the halo structure of $^{19}$,$^{17}$,$^{15}$C and $^{14}$B},
  {Phys. Rev. C}
   {57} 
   {(1998)}
   {2156}.
 \bibitem{Fang}
  {D.Q. Fang et al.},
  %title = {One-neutron halo structure in 15C},
  {Phys. Rev. C}
   {69} 
   {(2004)}
   {034613}.
 \bibitem{Freer}
  {M. Freer, H. Horiuchi, Y. Kanada-En'yo, D. Lee, and U.-G. Mei\ss{}ner},
  %title = {Microscopic clustering in light nuclei},
  {Rev. Mod. Phys.}
   {90} 
   {(2018)}
   {035004}.
 \bibitem{Hamamoto}
  {I. Hamamoto},
  %title = {Microscopic clustering in light nuclei},
  {Phys. Rev. C}
   {76} 
   {(2007)}
   {054319}.
   
\bibitem{TSuzukiI}
  {T. Suzuki, R. Fujimoto and T. Otsuka},
  %title = {Gamow-Teller transitions and magnetic properties of nuclei and shell evolution},
 {Phys. Rev. C}
   {67} 
   {(2003)}
   {044302}.

 \bibitem{TSuzukiII}
  {T. Suzuki and T. Otsuka},
  %title = {Exotic Magnetic properties in $^{17}C$},
  {Phys. Rev. C}
   {78} 
   {(2008)}
   {061301(R)}.

%\bibitem{VMU} T. Otsuka et al., Phys. Rev. Lett. 104 (2010) 012501.

 \bibitem{YSOX} C. Yuan et al., Phys. Rev. C 85 (2012) 064324.
 \bibitem{Kim} S. Kim et al., Phys. Lett. B 836 (2023) 137629.
   
\bibitem{chEFT_Epelbaum}
  {E. Epelbaum, H.-W. Hammer, and U.-G. Mei\ss{}ner},
  %title = {},
   {Rev. Mod. Phys.}
   {81} 
   {(2009)}
   {1773}.

 \bibitem{chEFT_Machleidt}
  {R. Machleidt and D. R. Entem},
  %title = {},
   {Phys. Rep.}
   {503} 
   {(2011)}
   {1}.
  

 \bibitem{Ekstrom}
   {A. Ekstr\"om et al.},
   %title = {Accurate nuclear radii and binding energies from a chiral interaction},
   {Phys. Rev. C}
   {91} 
   {(2015)}
   {051301(R)}.
   
\bibitem{VSoma3}
  {V. Som\`a, P. Navr\'atil, F. Raimondi, C. Barbieri and T. Duguet},
  %title = {Novel chiral Hamiltonian and observables in light and medium-mass nuclei},
  {Phys. Rev. C}
   {101} 
   {(2020)}
   {014318}.
   
\bibitem{Cecil}
  {F.E. Cecil, J.R. Shepard, R.E. Anderson, R.J. Peterson
and P. Kaczkowski},
  %title = {},
  {Nuc. Phys. A}
  {255} 
  {(1975)}
  {243}.

  

\bibitem{Goss1}
   {J.D. Goss et al.,}
   %{J.D. Goss, A.A. Rollefson, C.P. Browne, R.A. Blue, and H.R. Weller}
   {Phys. Rev. C}
   {8} 
   {(1973)}
   {514}.
   \bibitem{Goss2}
   {J.D. Goss et al.,}
   %{J.D. Goss, P.L. Jolivette, C.P. Browne, S.E. Darden, R.A. Blue, and H.R. Weller}
   {Phys. Rev. C}
   {12} 
   {(1975)}
   {1730}.
\bibitem{Murillo}
  {G. Murillo, S. Sen, and S.E. Darden},
  {Nucl. Phys. A}
  {579}
  {(1994)}
  {125}.
  \bibitem{Kay}
  {B. Kay et al.,}
  {Phys. Rev Lett.}
  {579}
  {(2022)}
  {125}.
  
  \bibitem{Truong}
   %{S. Truong et al.,}
   {S. Truong and H.T. Fortune}
   {Phys. Rev. C}
   {28} 
   {(1983)}
   {977}.
   
  \bibitem{Cappuzzello}
  {F. Cappuzzelo et al.},
  {Phys. Lett. B}
   {711} 
   {(2012)}
   {347}.
   
 \bibitem{UDatta}
  {U. Datta-Pramanik et al.},
  {Phys. Lett. B}
   {551} 
   {(2003)}
   {63}.

   \bibitem{Miyatake}
   {H. Miyatake et al.,}
   %{S. Truong and H.T. Fortune}
   {Phys. Rev. C}
   {67} 
   {(2003)}
   {014306}.

\bibitem{SauvanPLB} 
  {E. Sauvan et al.},
  {Phys. Lett. B}
   {491}
   {(2000)}
   {1}.
 
   
 \bibitem{Maddalena}
   {V. Maddalena et al.},
   %  {V. Maddalena and T. Aumann and D. Bazin and B. A. Brown and J. A. Caggiano and B. Davids and T. Glasmacher and P. G. Hansen and R. W. Ibbotson and A. Navin and B. V. Pritychenko and H. Scheit and B. M. Sherrill and M. Steiner and J. A. Tostevin and J. Yurkon},
   %title = {Single-neutron knockout reactions: Application to the spectroscopy of $^{16,17,19}$C},
   {Phys. Rev. C}
   {63}
   {(2001)}
   {024613}.
   

 \bibitem{Yamaguchi}
   {T. Yamaguchi et al.},
   %  { T.YamaguchiaT.ZhengabcA.OzawaaM.ChibaadR.KanungoaT.KatoaK.MorimotoaT.OhnishiaT.SudaaY.YamaguchiaeA.YoshidaaK.YoshidaaI.Tanihataa},
  %title = {Momentum distributions of 14C and 15C fragments from 16C breakup},
   {Nuc. Phys. A}
   {724}
   {(2003)}
   {3}.


% \bibitem{Wuosmaa}
%  {A.H. Wuosmaa et al.},
%  {A.H. Wuosmaa, B.B. Black, S. Baker, B.A. Brown, C.M. Deibel, P. Fallon, C.R. Hoffman, B.P. Kay, H.Y. Lee, J.C. Lightall, A.O. Macchiavelli, S.T. Marley, R.C. Pardo, K.E. Rehm, J.P. Schiffer, D.V. Shetty, and M. Wiedeking},
  %title = {$^{15}$C$(d,p)^{16}$C Reaction and Exotic Behavior in $^{16}C$},
%  {Phys. Rev. Lett.}
%   {105}
%   {(2010)}
%   {132501}.


  \bibitem{CATS}
  {S. Ottini-Hustache et al.},
%  {S. Ottini-Hustache and C. Mazur and F. Auger and A. Musumarra and N. Alamanos and B. Cahan and A. Gillibert and A. Lagoyannis and O. Maillard and E. Pollacco and J.L. Sida and M. Riallot},
  %title = {CATS, a low pressure multiwire proportionnal chamber for secondary beam tracking at GANIL},  {NIM A}
   {Nucl. Instrum. Meth. A}
   {431}  
   {(1999)}
   {476}.
  
\bibitem{TIARA}
  {M. Labiche et al.},
%  {M. Labiche and W.N. Catford and R.C. Lemmon and C.N. Timis and R. Chapman and N.A. Orr and B. Fern\'{a}ndez-Dom\'{i}guez and G. Moores and N.L. Achouri and  N. Amzal and S. Appleton and N.I. Ashwood and T.D. Baldwin and M. Burns and L. Caballero and J. Cacitti and J.M. Casadjian and M. Chartier and  N. Curtis and  K. Faiz and G. de France and M. Freer and J.M. Gautier and W. Gelletly and G. Iltis and B. Lecornu and X. Liang and C. Marry and Y. Merrer and L. Olivier and S.D. Pain and V.F.E. Pucknell and B. Raine and M. Rejmund and B. Rubio and F. Saillant and H. Savajols and O. Sorlin and K. Spohr and Ch. Theisen and G. Voltolini and D.D. Warner},
  %title = {TIARA: A large solid angle silicon array for direct reaction studies with radioactive beams},  {NIM A}   {614} 
   {Nucl. Instrum. Meth. A}
   {614} 
   {(2010)}
   {439}.


   \bibitem{MUST2}
  {E. Pollacco et al.},
%  {E. Pollacco and D. Beaumel and P. Roussel-Chomaz and E. Atkin and P. Baron and J. P. Baronick and E. Becheva and Y. Blumenfeld and A. Boujrad and A. Drouart and F. Druillole and P. Edelbruck and M. Gelin and A. Gillibert and Ch. Houarner and V. Lapoux and L. Lavergne and G. Leberthe and L. Leterrier and V. Le Ven and F. Lugiez and L. Nalpas and L. Olivier and B. Paul and B. Raine and A. Richard and M. Rouger and F. Saillant and F. Skaza and M. Tripon and M. Vilmay and E. Wanlin and M. Wittwer},
  %title = {MUST2: A new generation array for direct reaction studies},
  {Eur. Phys. J. A}
   {25}  
   {(2005)}
   {287}.
   

\bibitem{CHARISSA}
  {N. I. Ashwood et al.},
  {Phys. Rev. C}
   {70}  
   {(2004)}
   {024608}.

   \bibitem{EXOGAM}
  {J. Simpson et al.},
%  {J. Simpson and F. Azaiez and G. DeFrance and J. Fouan and J. Gerland R. Julin and W. Korten and P. J. Nolan and B. Nyak盧錦 and G. Sletten and P. M. Walker},
  %title = {The EXOGAM array},
  {Acta Phys. Hung., New Series, Heavy Ion Physics},
   {11}  
   {(2000)}
   {159}.
   
 \bibitem{ENSDF}
  {ENSDF},  
  {https://www.nndc.bnl.gov/ensdf/}
  
    \bibitem{BlCrom}
  {B. Le Crom et al.},
%  {B. Le Crom, M.Assié, Y.Blumenfeld, J.Guillot, H.Sagawa, T.Suzukic, .Honmab, N.L.Achouri, B.Bastin,R.Borcea, W.N.Catford,E.Clément, L.Cáceres, M.Caamaño, A.Corsi, G.De France,F.Delaunay,N.De Séréville,B.Fernandez-Dominguez,M.Fisichella,S.Franchoo,A.Georgiadou,J.Gibelin,A.Gillibert,F.Hammache,O.Kamalou,A.Knapton,V.Lapoux,S.Leblond,A.O.Macchiavelli,F.M.Marqués,A.Matta, L.Ménager,P.Morfouace,N.A.Orr,J.Pancin,X.Pereira-Lopez,L.Perrot,J.Piot,E.Pollacco,D.Ramos,T.Roger,F.Rotaru,A.M.Sánchez-Benítez,M.Sénoville,O.Sorlin,M.Stanoiu,I.Stefan,C.Stodel,D.Suzuki,J.-C.Thomas,M.Vandebrouck},
  %title = {Neutron-proton pairing in the N=Z radioactive fp-shell nuclei 56Ni and 52Fe probed by pair transfer},  {PlB}
   {Phys. Lett. B}
   {829}  
   {(2022)}
   {137057}.

    \bibitem{XPereiraPLB}
     {X. Pereira-L\'opez et al.},
  %{https://minerva.usc.es/xmlui/handle/10347/15161}
     {Phys. Lett. B}
     {811}
     {(2020)},
     {135939}.
     
    \bibitem{XPereiraPhD}
  {X. Pereira-L\'opez, ``Study of transfer reactions induced by a $^{16}$C beam'', PhD Thesis (2016)},
  %{https://minerva.usc.es/xmlui/handle/10347/15161}
  {http://hal.in2p3.fr/tel-01522695}

   \bibitem{nptool}
  { A. Matta, P. Morfouace, N. de S\'{e}r\'{e}ville, F. Flavigny, M. Labiche and R. Shearman},
  %title = {NPTOOL},  {-}
   {J. Phys. G}
   {43}  
   {(2016)}
   {045113}.  

  \bibitem{FRESCO}
   {I.J. Thompson},
   {Comp. Phys. Rep.}
   {7}
   {(1988)}
   {167}.
   
   {http://www.fresco.org.uk/}
 \bibitem{Haixia}
   {Haixia An and Chonghai Cai},
   %title = {Global deuteron optical model potential for the energy range up to 183 MeV},
   {Phys. Rev. C}
   {73}
   {(2006)}
   {054605}.
 \bibitem{Pang}
   {D.Y. Pang et al.},
   %title = {Neutron spectroscopic factors from transfer reactions},
   {Phys. Rev. C}
   {79}
   {(2009)}
   {024615}.
 \bibitem{Brida}
   {I. Brida, S.C. Pieper, and R.B. Wiringa},
  %title = {Neutron spectroscopic factors from transfer reactions},
   {Phys. Rev. C}
   {84}
   {(2011)}
   {024319}.
   
\bibitem{MF-sumrule}
  {M. H. Macfarlane and J.B. French}
   {Rev. Mod. Phys.}
   {32} 
   {(1960)}
   {567}.
    \bibitem{BAYE}
     {D. Baye and N. Timofeyuk},
     %title = {Vertex constants and the problem of the nucleon-nucleon potential in the generator coordinate method},
     {Phys. Lett. B}
     {293}
     {(1992)}
     {13}.
   \bibitem{BFD21Al}
     {N. Timofeyuk et al.},
     %title = {Core excitations and narrow states beyond the proton dripline: The exotic nucleus 21Al},
     {Phys. Rev. C}
     {86}
     {(2012)}
     {034305}.
   \bibitem{BFD25P}
     {B. Fern\'andez-Dom\'inguez et al.},
     %title = {Core excitations and narrow states beyond the proton dripline: The exotic nucleus 21Al},
     {Phys. Rev. C}
     {91}
     {(2015)}
     {024307}.
   
% \bibitem{LeeSF}
%   {J. Lee, M. B. Tsang and W. G. Lynch},
%   %title = {Neutron spectroscopic factors from transfer reactions},
%   {Phys. Rev. C}
%   {75}
%   {(2007)}
     %   {064320}.

        
   \bibitem{VSoma1}
  {V. Som\`a, T. Duguet, and  C. Barbieri},
  %title = {\texit{Ab initio} self-consistent Gorkov-Green's function calculations of semimagic nuclei: Formalism at second order with a two-nucleon interaction},
  {Phys. Rev. C}
  {84} 
  {(2011)}
  {064317}.
  
\bibitem{VSoma2}
  {V. Som\`a, C. Barbieri and T. Duguet},
%  %title = {\textit{Ab initio} self-consistent Gorkov-Green's function calculations of semi-magic nuclei: Numerical implementation at second order with a two-nucleon interaction},
  {Phys. Rev. C}
   {89} 
   {(2014)}
   {024323}.
   
    \bibitem{Lapoux}
  {V. Lapoux et al.},
%  %title = {\textit{Ab initio} self-consistent Gorkov-Green's function calculations of semi-magic nuclei: Numerical implementation at second order with a two-nucleon interaction},
  {Phys. Rev. Lett.}
   {117} 
   {(2016)}
   {052501}.

  \bibitem{VSoma4}
  {V. Som\`a, C. Barbieri, T. Duguet and P. Navr\'atil},
%  %title = {\textit{Ab initio} self-consistent Gorkov-Green's function calculations of semi-magic nuclei: Numerical implementation at second order with a two-nucleon interaction},
  {Eur. Phys. J. A}
   {57} 
   {(2021)}
   {135}.

   \bibitem{Mairle}
  {G. Mairle et al., }
      %{G. Mairle, K.T. Knöpfle, P.Doll, H. Breuer, and G.J. Wagner}
   {Nucl. Phys. A}
   {280} 
   {(1977)}
   {97}.
   
 \bibitem{MStanoiu-N=14Cchain}
  {M. Stanoiu et al.},
  %  {M. Stanoiu and D. Sohler and O. Sorlin and F. Azaiez and Zs. Dombr\'{a}di and B. A. Brown and M. Belleguic and C. Borcea and C. Bourgeois and Z. Dlouh\'y and Z. Elekes and Zs. F\"ul\"op and S. Gr\'evy and D. Guillemaud-Mueller and F. Ibrahim and A. Kerek and A. Krasznahorkay and M. Lewitowicz and S. M. Lukyanov and S. Mandal and J. Mr\'{a}zek and F. Negoita and Yu.-E. Penionzhkevich and Zs. Podoly\'{a}k and P. Roussel-Chomaz and M. G. Saint-Laurent and H. Savajols and G. Sletten and J. Tim\'{a}r and C. Timis and A. Yamamoto},
  %title = {Disappearance of the N=14 shell gap in the carbon isotopic chain},
  {Phys. Rev. C}
   {78} 
   {(2008)}
   {034315}.
   
%\bibitem{Ajzenberg}
%      {F. Ajzenberg-Selove},
%   {Nucl. Phys. A}
%   {523} 
%   {(1991)}
%   {1}.
%\bibitem{MF-sumrule}
%  {M. H. Macfarlane and J.B. French}
%   {Rev. Mod. Phys.}
%   {32} 
%   {(1960)}
%   {567}.
%\bibitem{Garret}
%  {J.D. Garret et al.,}
%   %{J.D. Garret, F. Ajzenberg-Selove, And H.G. Bingham}
%   {Phys. Rev. C}
%   {10} 
%   {(1973)}
%   {1730}.
   


% \bibitem{Fortunesd}
%  {H.T. Fortune et al., }
%      %{H.T. Fortune, M.E. Cobern and G.E. Moore}
%   {Phys. Rev. C}
%   {17} 
%   {(1978)}
%   {888}.


  
   \bibitem{barbieri22}
  {C. Barbieri, T. Duguet and V. Som\`a},
  %title = {Neutron spectroscopic factors from transfer reactions},
  {Phys. Rev. C}
   {105}
   {(2022)}
   {044330}.

%\bibitem{PAMD}
%  {J. A. Lay et al.},
%%  {J. A. Lay and A. M. Moro and J. M. Arias and Y. Kanada-En'yo},
  %title = {Semi-microscopic folding model for the description of two-body halo nuclei},
%  {Phys. Rev. C}
%   {89}
%   {(2014)}
%   {014333}.

%\bibitem{WarburtonBrown}
%  {E. K. Warburton and B. A. Brown},
%  %title = {Effective interactions for the $0p1s0d$ nuclear shell-model space},
%  {Phys. Rev. C}
%   {46}
%   {(1992)}
%   {923}.

%\bibitem{Baranger}
%M. Baranger, Nucl. Phys. A 149 (1970) 225.

%\bibitem{Signoracci}
%  {A. Signoracci and B.A. Brown},
%  %title = {Comment on ``Reduction of the Spin-Orbit Splittings at the N = 28 Shell Closure''},
%  {Phys. Rev. Lett.}
%   {99}
%   {(2007)}
%   {099201}.

%\bibitem{MCAS}
%K. Amos et al., Nucl. Phys. A 879 (2012) 132.

%\bibitem{Be15}
%  {J. Snyder et al.},
%  {J. Snyder and T. Baumann and G. Christian and R. A. Haring-Kaye and P. A. DeYoung and Z. Kohley and B. Luther and M. Mosby and S. Mosby and A. Simon and J. K. Smith and A. Spyrou and S. Stephenson and M. Thoennessen},
%  %title = {First observation of $^{15}$Be},
%  {Phys. Rev. C}
%   {88}  
%   {(2013)}
 %  {031303}.
  
%  \bibitem{ELevelsA19}
%  {D.R. Tilley et al.},
%%  {D.R. Tilley and H.R. Weller and C.M. Cheves and R.M. Chasteler},
%  %title = {Energy Levels of Light Nuclei $A=19$},
%  {Nucl. Phys. A}
%   {595}
%   {(1995)}
%   {1}
%   {and references therein}.




   
   
\end{thebibliography}

\end{document}